\documentclass[12pt]{article}

\usepackage{newtxtext} 
\usepackage[top=2.5cm, bottom=2.5cm, left=3cm, right=3cm]{geometry} 
\usepackage[english]{babel}
\usepackage{setspace} 
\usepackage[parfill]{parskip} 
\usepackage[english]{babel}

\usepackage{amsmath}
\usepackage{graphicx}
\usepackage[colorlinks=true, allcolors=blue]{hyperref}
\usepackage{multirow}
\usepackage{caption}
\usepackage{placeins}
\usepackage{rotating}
\usepackage{apacite}
\usepackage{subfigure}
\usepackage{booktabs}
\usepackage{threeparttable}
\usepackage{natbib}
\usepackage{hyperref}

\title{Effectiveness of Rent Controls: Evidence from Spain}
\author{Luis Pérez García\thanks{I would like to express my sincere gratitude to the \textit{Consellería de Hacienda, Economía y Administración
Pública} for their generous financial support under the grant HIUNPU/2024/02, which made my studies and this work possible.}}
\date{July 2025}

\begin{document}
\maketitle

\section*{Abstract}

Growing concerns about housing affordability have prompted the adoption of rent control policies and renewed debates over their effectiveness. This paper provides the first empirical evaluation of the 2024 rent control policy implemented in Catalonia under Spain’s new national housing law. To identify the causal effect of the policy on the rental market, I use municipality-level administrative data and implement several difference-in-differences strategies and event study designs. The results point to a reduction in tenancy agreements and a less robust decrease in rental price growth. While the findings highlight important short-term consequences of rent control, they also underscore the need for caution due to data limitations and limited robustness in some estimates.

\vspace{1cm}

\textbf{Keywords:} housing, rental prices, rent controls, housing affordability.

\clearpage
\section{Introduction}

Housing affordability has recently become one of the most pressing concerns in cities throughout Europe, as rental prices continue to outpace income growth for many households. In response, policymakers have increasingly turned to rent controls as a tool to stabilize rents and protect tenants from volatile market dynamics. Major cities such as Berlin and Paris exemplify this growing trend.

More recently, the regional government of Catalonia (Spain) has implemented two rent control measures with similar objectives but slightly different designs. The first was introduced between 2020 and 2022 and applied to municipalities with more than 20,000 inhabitants that were classified as having a tight rental market, although the criteria for this classification were not entirely transparent. The second policy came into effect in March 2024 and remains in force. It established clearer conditions for identifying stressed housing markets and removed the population threshold, significantly expanding the number of covered municipalities. Despite these differences in scope and targeting, both regulations shared a common core: in treated areas, new rental contracts were prohibited from exceeding the lower of either a reference price index or the rent paid under the previous lease for the same unit.

This work evaluates the effectiveness of the 2024 rent control policy implemented in Catalonia, focusing on its ability to control rental prices — its primary policy goal— while assessing potential unintended consequences on rental housing supply, a common concern in the rent control literature. Understanding these effects is especially relevant in the broader context of worsening housing affordability in Spain and other European countries, where policymakers are increasingly turning to price regulation as a tool to stabilize rental markets. I analyze two key outcomes aligned with the policy’s objectives and criticisms: the growth rate of rental prices and the number of tenancy agreements signed, which serves as a proxy for rental housing supply. The analysis relies on newly released administrative data from the \textit{Generalitat de Catalunya} and applies a difference-in-differences framework, complemented by event study designs and other specifications to improve comparability of treated and control units - such as propensity score matching and synthetic control methods. 

The results from the empirical analysis suggest that the 2024 rent control policy in Catalonia was associated with a decline in both rent growth and the number of tenancy agreements signed. However, event study estimates reveal evidence of diverging pre-treatment trends between treated and control municipalities, calling into question the validity of the parallel trends assumption required for causal interpretation. When alternative specifications that aim to alleviate this problem are applied, the estimated effect on rent prices largely disappears, while the reduction in rental supply remains robust. Taken together, these findings point to a potentially negative impact of the policy on rental housing availability, although we should be cautious when drawing causal conclusions due to data limitations and sensitivity to model specification.

The empirical evidence on rent controls is quite recent, scarce and usually yields mixed results — especially when it comes to whether these policies successfully keep rents from rising in the long run without reducing the supply of rental housing. For the US case, early work by \cite{sims2007} and \cite{autor2014} focused on the aftermath of rent control deregulation in Massachusetts during the 1990s. These studies found that rents rose more sharply in areas previously covered by controls, but some of this effect may reflect neighborhood gentrification processes. In the case of San Francisco, \cite{diamond2019} studied a 1994 reform that expanded rent control to previously exempt buildings. Their findings suggest that the policy slowed down the displacement of low-income tenants, indicating potential protective effects for vulnerable households. \cite{mense2019}, \cite{mense2023} and \cite{breidenbach2022} analyzed Germany’s 2015 federal rental regulation, which introduced caps in municipalities with strained housing markets. While these studies find some deceleration in rental price growth, the effects appear short-lived, with limited persistence beyond the first year.

Lastly, two recent studies evaluate the effects of the first rent control policy implemented in Catalonia, both using microdata from INCASOL, which contains the universe of registered rental contracts in the region between 2016 and 2022. \cite{monras2022} estimate that the policy led to a 5\% reduction in rental prices and interpret this outcome through a model with search frictions. However, they also document a decline in rental supply, particularly among high-priced units. In contrast, \cite{jofre2023} report a similar reduction in rents but find no evidence of impact on the supply of rental housing.

This paper contributes to the existing literature by providing the first empirical evaluation of the rent control policy implemented in Catalonia in 2024 under the framework of the \textit{Ley por el Derecho a la Vivienda}. Using the most recent released administrative data and a difference-in-differences approach, it offers early evidence on how this new national legal framework affects rental prices and housing supply at the municipal level. In doing so, it extends the literature on rent regulation to a new institutional setting and informs ongoing debates about the effectiveness and design of price control mechanisms in the housing market.

\section{Background}

Spanish Housing market is dominated by homeowning. In 2023, approximately 75\% of households lived in owner-occupied housing, while only 25\% were renters. Most rental agreements take place in the private market, with social housing and informal arrangements accounting for a small share. Despite these low figures, the proportion of market-rate renters has grown steadily in recent years, rising from around 10\% in 2004 to 15\% in 2023, as homeownership has gradually declined.\footnote{Source: \textit{Encuesta de Condiciones de Vida}, INE.}

Nevertheless, housing affordability has consistently ranked among the most pressing concerns for Spanish households in recent years, according to successive public opinion surveys.\footnote{Source: \textit{Barómetros from Centro de Investigaciones Sociológicas}, CIS.} In fact, Spain ranks among the least affordable countries in the OECD when it comes to housing costs. Around 20\% of tenants spend over 40\% of their income on rent, placing them in a situation of severe financial stress.\footnote{Source: OECD Affordable Housing Database.}

As an answer to the housing affordability problems, in May 2023 was enacted the \textit{Ley por el Derecho a la Vivienda}, the first national housing law in Spain to include a legal framework for rent regulation. A central feature of this law is the concept of “stressed residential market areas” (\textit{zonas de mercado residencial tensionado}), which are municipalities where regional governments can set a cap on rental prices. A municipality can be classified as stressed if either of the following conditions is met: (i) the average housing cost exceeds 30\% of average household income, or (ii) housing costs have increased by more than three percentage points above inflation over the previous five years.

The rent control introduced by this law vary depending on the type of landlord\footnote{Large landlords are defined as those owning ten or more residential rental properties, or more than 1,500 m\textsuperscript{2} of built residential floor area.}. Large landlords are subject to stricter regulation: in stressed areas, rents for new contracts may not exceed the lower of either the amount charged in the previous lease for the same unit or the value determined by a reference price index. In contrast, small landlords are not subject to the reference index, unless the unit is being rented for the first time or has remained vacant for a long period, so only the previous price limit matters.

Catalonia was the first region to implement the rent control provisions of the new national law, designating 140 municipalities as stressed areas on March 16\textsuperscript{th} 2024, more than one year after the enactment, followed by an additional 131 municipalities on October 10\textsuperscript{th}, 2024. This marked the second major intervention in rent regulation by the Catalan government in recent years. Between September 2020 and March 2022, Catalonia had already conducted an earlier rent control experiment under a regional law. That policy introduced similar price caps in 61 municipalities with more than 20,000 inhabitants and rapidly rising rents, all of which were again included in the first wave of municipalities subject to the 2024 regulation.

\section{Data}

I use publicly available data from the regional government of Catalonia on average rental prices and the number of tenancy agreements at the municipality and quarterly level. The dataset covers the period from 2005 to 2024 (inclusive), although most of the analysis focuses on data from 2022 onward, since several treated municipalities were already subject to a similar rent control policy between in 2020 and 2021. A central limitation of this dataset is that the only available price metric is the average rental price per municipality, which is not adjusted for dwelling size. As a result, differences in average prices across municipalities may reflect variations in average dwelling size rather than underlying price differences per square meter. To mitigate this issue and reduce the impact of composition effects, I use the growth rate of average rental prices at the municipality level, rather than price levels in absolute terms. In addition, I control for other covariates at the municipality level that may be correlated with the average size of houses. Nevertheless, the possibility of composition effects cannot be entirely ruled out and should be kept in mind when interpreting the results.

One important limitation concerns the second wave of treated municipalities, which were designated as stressed areas in October 2024. Since the available data ends in 2024, only a single post-treatment quarter is observed for these municipalities. Including them in the analysis at this stage could introduce excessive noise and bias due to insufficient post-treatment data. Therefore, I exclude the second wave from the empirical analysis and focus exclusively on the first wave of treated municipalities, for which a longer post-treatment period is available. Once new data is released, the analysis will be extended to incorporate the second wave of treated municipalities.

Another important data limitation is the presence of missing observations for rental prices, particularly among control municipalities. While rental data is available for all treated municipalities — with only seven experiencing missing values in one or more quarters — the situation is more limited in the control group. Specifically, rent data are available for only 173 out of the 680 control municipalities, and even among these, many display missing data in some quarters. This substantially restricts the pool of municipalities that can be used to construct a credible counterfactual. Moreover, the distribution of missing data between the treatment and control groups appears clearly non-random. While this may not necessarily bias the results, it raises concerns if the missingness is correlated with the outcome variables or other relevant characteristics.

I complement the data on rent prices and tenancy agreements at the municipality level with other data sources to construct control variables. Specifically, I use administrative data on average income at the municipality level from the Spanish Tax Agency (AEAT) to create a proxy for the rent-to-income ratio. This ratio is one of the criteria used by the regional government to declare stressed residential market areas. An alternative identification strategy could have been a regression discontinuity design (RDD) around the 30\% threshold used in this classification. However, this approach requires access to the exact variables and methodology used by the authority to compute average rental costs and household income at the municipality level. Since the government has not disclosed the precise data sources or calculation method, it is not possible to replicate the assignment rule with sufficient accuracy. As a result, RDD is not a viable strategy in this context\footnote{In fact, 53 out of the 140 treated municipalities fall below the 30\% threshold based on our proxy, and 25 out of the 173 control municipalities exceed it.}. Finally, I include data on population size from the Census, as well as local population inflows and outflows from the \textit{Encuesta de Variaciones Residenciales} (INE).

Table~\ref{tab:descriptive_stats} presents summary statistics for the treatment and control groups. The first panel displays key variables both before and after the implementation of the rent control policy. Although average rental prices differ between the two groups, pre-treatment rent growth rates are nearly identical. This suggests that potential composition effects present in average prices due to differences in dwelling size are likely being mitigated by focusing on rent growth, provided that this size has not changed significantly during the one-year treatment period. In the post-treatment period, however, rent growth is considerably lower in treated municipalities, offering preliminary evidence that the policy may have pressured rental prices downward. A similar dynamic is observed in the number of tenancy agreements: while the volume remain relatively stable in the control group across periods, it declines in the treated group following the introduction of the policy, potentially signaling a reduction in rental supply.

Despite these patterns, the second panel of the table highlights systematic differences between the treatment and control groups along several key dimensions. On average, control municipalities are smaller, less populated, and exhibit lower income levels. They also display lower rent-to-income ratios and different migration inflow patterns, indicating a less stressed housing market overall. These structural differences underscore the limitations of using the control group as a direct counterfactual for the treated municipalities and motivate the need for a difference-in-differences design with appropriate controls.

\begin{table}[ht]
\centering
\caption{Descriptive statistics for treated and control municipalities}
\label{tab:descriptive_stats}
\begin{tabular}{lrrr}
  \toprule
Variable & Control & Treated \\ 
  \midrule
Rent Price Pre & 506.158 & 681.049 \\ 
  Rent Price Post & 577.825 & 760.140 \\ 
  Yearly Rent Growth (2019 - 2024) & 0.038 & 0.039 \\ 
  Yearly Rent Growth (2022 - 2024) & 0.070 & 0.072 \\ 
  Rent Growth Post & 0.045 & -0.005 \\ 
  Tenancy Agreements Pre & 30.521 & 42.443 \\ 
  Tenancy Agreements Post & 30.500 & 34.993 \\ 
  \midrule
  Rent Price/Net Income & 0.251 & 0.305 \\ 
  Yearly Gross Income & 30,837 & 34,518 \\ 
  Population & 3,393 & 45,536 \\ 
  Net Inflows from Natives & 16.381 & -129.229 \\ 
  Net Inflows from Foreigners & 15.278 & 223.943 \\
  \midrule
  Number of Municipalities & 680 & 140 \\ 
  Municipalities with Data on Rents & 173 & 140 \\
  Municipalities without Missing Data & 27 & 133 \\
  \bottomrule
\end{tabular}
\end{table}

\section{Empirical Strategy}

To estimate the causal effect of the rent control, I employ a difference-in-differences framework (DiD). The baseline specification is given by the following regression equation:

\begin{equation}
    Y_{it} = \alpha + \beta (Treatment_{i} \times Post_{t}) + \delta X_{it} + \gamma_{i} + \gamma_{t} + \epsilon_{it}
\end{equation}

where $Y_{it}$ denotes the outcome variable of interest for municipality $i$ in quarter $t$, namely, the growth rate of rental prices or the number of tenancy agreements per 10,000 inhabitants.

The main explanatory variable is the interaction term $Treatment_{i} \times Post_{t}$, where $Treatment_{i}$ indicates whether municipality $i$ was subject to rent control under the first wave of implementation in March 2024, and $Post_{t}$ is a time indicator equal to one in quarters when the policy is in effect. Since the first wave of municipalities were designated as stressed on March 16\textsuperscript{th}, 2024, I consider the second quarter of 2024 as the start of the treatment period. The coefficient $\beta$ captures the average difference in rent growth (or the number of tenancy agreements) between treated and control municipalities after the policy was introduced. Standard errors are clustered at the municipality level.

In some specifications, I include municipality fixed effects and time fixed effects ($\gamma_{i}$ and $\gamma_{t}$), along with additional covariates that may be correlated with housing outcomes. Specifically, the control vector $X_{it}$ includes: (i) total population, to account for municipality size; (ii) net migration flows of natives and foreigners, to control for shifts in population composition; (iii) the rent-to-income ratio, which proxies for housing market stress; and (iv) the number of tenancy agreements in the pre-treatment period, when rent growth is the dependent variable (and vice versa when modeling tenancy agreements).

To complement the baseline DiD analysis, I implement an event study specification that allows for a more flexible estimation of the dynamic effects of rent control on the residential market. This approach also enables a visual and statistical assessment of pre-treatment trends, helping to detect potential anticipation effects or violations of the parallel trends assumption. The estimated specification is as follows:

\begin{equation}
    Y_{it} = \alpha + \sum_{t \neq 2024Q1} \beta_{t} (Treatment_{i} \times \gamma_{t}) + \delta X_{it} + \gamma_{i} + \gamma_{t} + \epsilon_{it}
\end{equation}

As in the DiD regression, I include covariates as well as municipality and time fixed effects. The coefficients of interest are the $\beta_{t}$ parameters, which correspond to the interaction between the treatment indicator and a set of quarter dummies. I normalize the estimates relative to the first quarter of 2024, the last period before the policy was implemented.

As we will see in the following section, there seems to be enough evidence of a lack of parallel trends, which could lead to a biased analysis using simple DiD. Hence, I complement that approach with other DiD techniques whose main focus is to achieve parallel trends by weighting differently each municipality in the control pool to compare the treated units to the most similar one in the control units. Specifically, I use two different frameworks: propensity score matching DiD and synthetic DiD. The PSM-DiD estimator constructs a matched sample of control municipalities that are observationally similar to the treated ones based on a set of pre-treatment covariates. The DiD model is then applied within this matched sample, aiming to reduce bias from observable differences. The Synthetic DiD method, following the approach proposed by \cite{arkhangelsky2021}, combines the strengths of synthetic control and DiD by constructing a weighted combination of control units whose trend in the outcome variable is parallel the treated group’s. This method adjusts for both observed and unobserved confounders, provided they are reflected in pre-treatment trends.

\section{Results}

\subsection{Baseline Results}

Table~\ref{tab:did_table_cluster} reports the results from estimating Equation~(1). In column (1), rent growth is regressed on the DiD interaction term without controls. Column (2) adds the main covariates described in the data section. Column (3) presents the preferred specification, which includes both municipality and time fixed effects in addition to the control variables. Columns (4) through (6) replicate the same sequence of specifications using the number of tenancy agreements per 10,000 inhabitants as the dependent variable.

The results in columns (1) to (3) suggest that the rent control policy reduced the growth rate of average rental prices, although the effect is statistically significant only in the specifications that include covariates. In the preferred specification (column 3), the estimated coefficient indicates that rent growth was approximately 3 percentage points lower in treated municipalities relative to the control group. To contextualize this effect, average annual rent growth in the years prior to treatment was 7.2\%. Moreover, the average rent in treated municipalities before the policy was 681.05 euros per month. Applying a 3\% to this baseline implies a drop of roughly 20.5 euros per month, or about 246 euros per year. This reduction in housing costs is economically meaningful, particularly when compared to the average net household income in Catalonia, which was 27,645 euros in 2022, the last available year.

Turning to columns (4) to (6), the results of all specifications suggest a significant reduction of around 6 tenancy agreements per 10,000 inhabitants in treated municipalities. Given a pre-treatment benchmark of 42 tenancy agreements per 10,000 inhabitants, this represents a decline of around 13\% in tenancy agreements signed. While this drop likely reflects a contraction in the supply of rental housing, it may also capture other dynamics. For instance, some units might continue to be rented informally, be shifted from the rental to the ownership market or be converted into short-term tourist rentals such as Airbnbs. The informal rental explanation seems less likely in this context, given the strong penalties that were established as enforcement measures associated with the 2023 law. That is, tenants could credibly threaten to report non-compliance, which likely deters informal rental arrangements that violate the regulation. A more plausible channel is that landlords have either opted to sell their units or moved them to the more flexible and potentially more profitable tourist rental market, which is not subject to the same price regulations. While this hypothesis could have further reaching implications for housing supply and affordability, testing it lies beyond the scope of this paper and is left for future research.

\begin{figure}[htbp]
  \centering
  
  \includegraphics[width=0.9\textwidth]{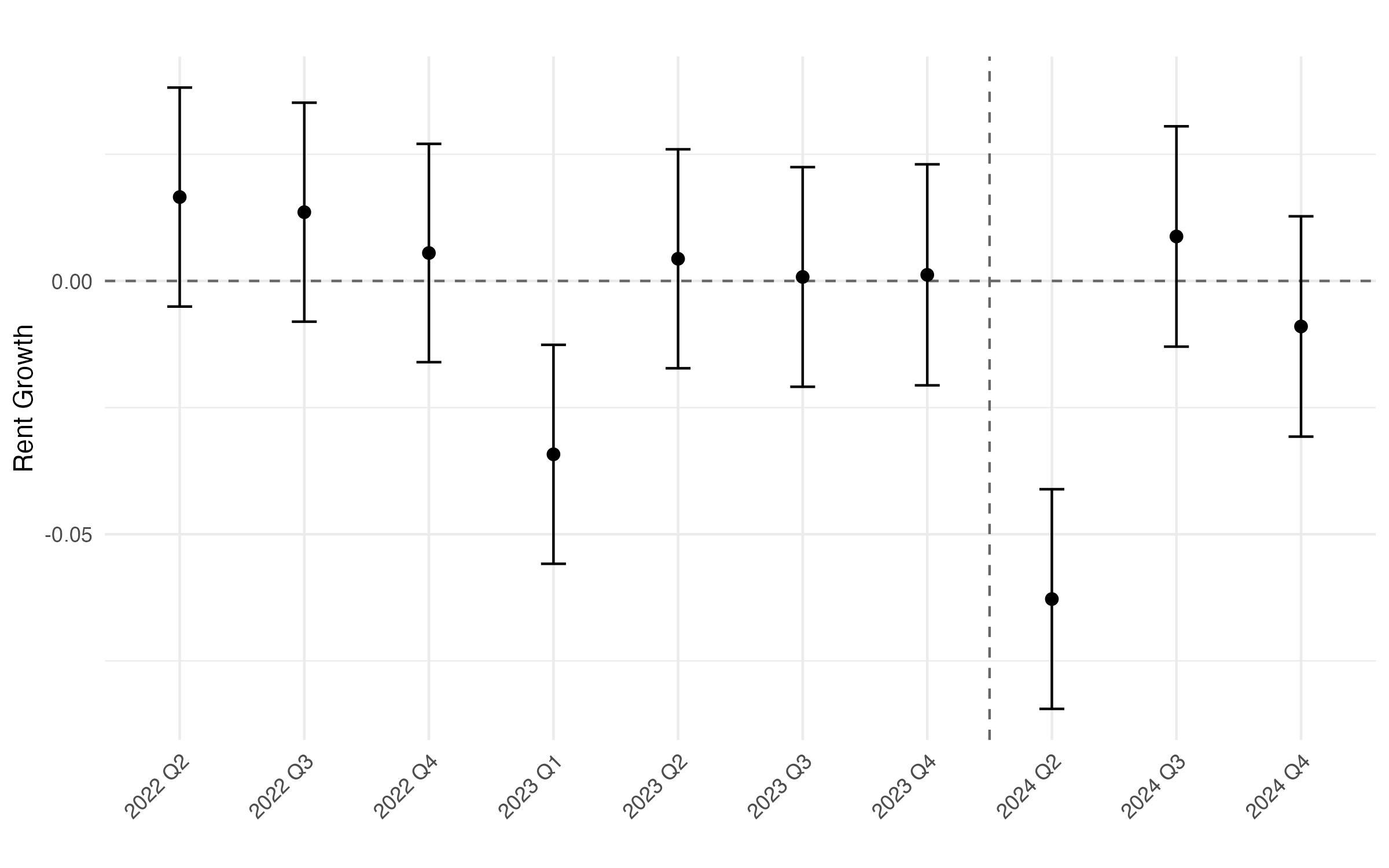}
  \caption*{(a) Rent Growth}
  
  \vspace{1em}
  
  \includegraphics[width=0.9\textwidth]{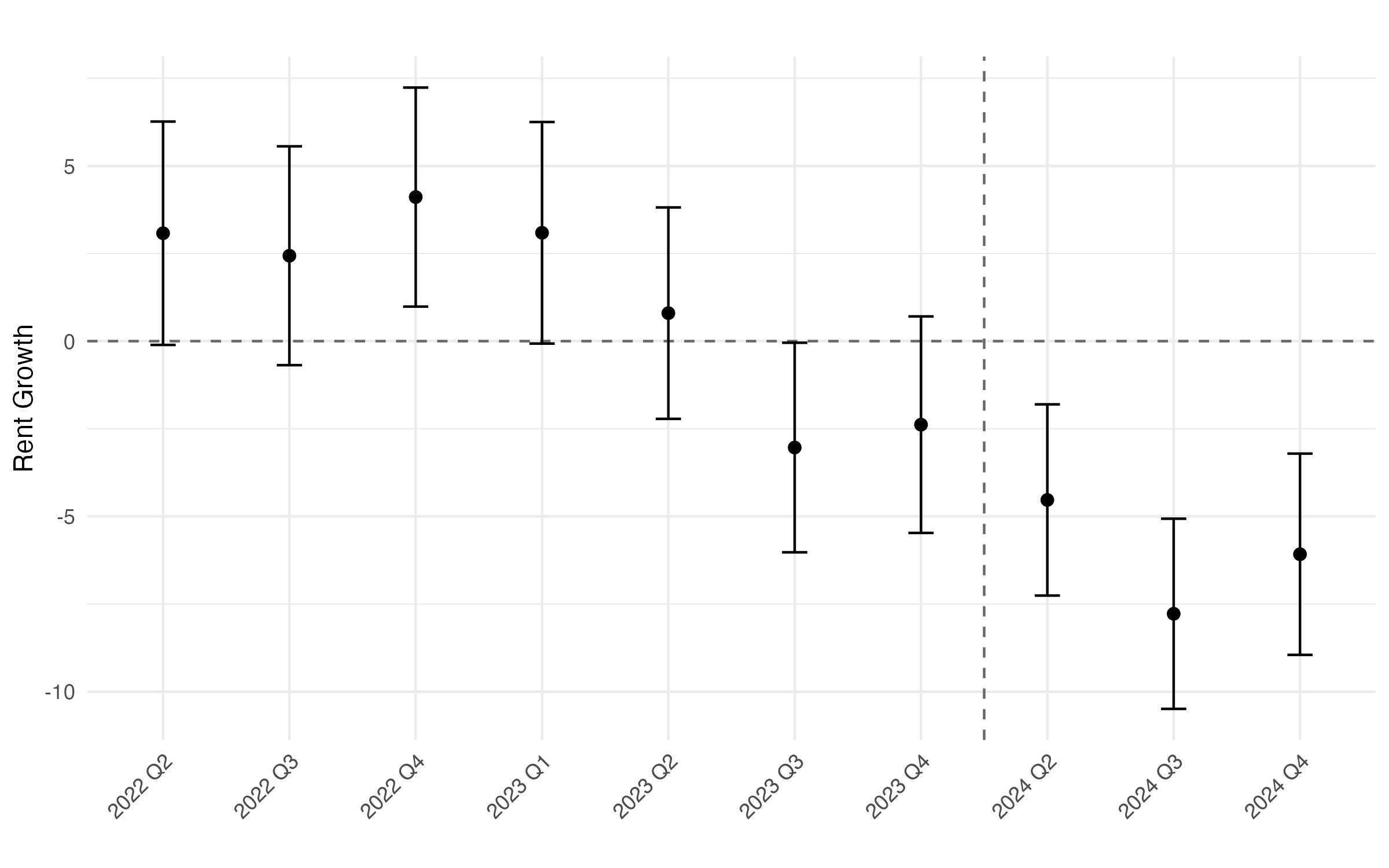}
  \caption*{(b) Tenancy Agreements per 10,000 Inhabitants}
  
  \caption{Event Study Results}
  \label{fig:event_study}
\end{figure}

The results from the event study regressions based on Equation~(2) are reported in Figure~\ref{fig:event_study}, which plots the estimated coefficients on the interaction between the treatment indicator and quarterly dummies, along with their 95\% confidence intervals. Panel~(a) shows the results for rent price growth, while Panel~(b) displays the estimates for tenancy agreements per 10,000 inhabitants.

Panel~(a) offers two key insights. First, it confirms that rent growth declined immediately after the implementation of the policy, but only for a single quarter. The estimated drop in that period is approximately 6.3 percentage points, suggesting that the 3 percentage point effect captured in the DiD specification represents an average across post-treatment periods. It is worth noting that even a temporary decline in the growth rate can lead to a lasting reduction in rent levels. Second, the pre-treatment dynamics indicate a deviation from the parallel trends assumption: the interaction term for the first quarter of 2023 is statistically significant. Moreover, a joint significance test confirms that the pre-treatment coefficients are jointly different from zero (p-value = 0.0223), as are the post-treatment coefficients (p-value = 1.15e-07).

Panel~(b) shows a consistent reduction in the number of tenancy agreements signed in treated municipalities, ranging from 4.5 to 7.8 fewer contracts per 10,000 inhabitants per quarter. However, the divergence in pre-treatment trends is even more pronounced than in Panel~(a). Once again, a joint test confirms that both the pre-trend (p-value = 8.28e-05) and the post-trend (p-value = 1.74e-10) are significantly different from zero.

Taken together, the event study results corroborate the findings from the DiD regressions, indicating that the rent control policy was associated with lower rent growth and a decline in rental market activity. However, we should be cautious when interpreting these results as causal since there are evidence against the parallel trends assumption, which is fundamental for DiD methods.

\begin{table}[htbp]
\centering
\caption{Baseline DiD Regressions}
\label{tab:did_table_cluster}
\begin{threeparttable}
\begin{tabular}{lcccccc}
\toprule
 & \multicolumn{3}{c}{Rent Growth} & \multicolumn{3}{c}{Tenancy Agreements} \\
 & (1) & (2) & (3) & (4) & (5) & (6) \\
\midrule
Treatment $\times$ Post    & -0.0172        & -0.0218$^{**}$              & -0.0298$^{**}$          & -6.380$^{***}$ & -5.995$^{***}$ & -6.586$^{***}$\\   
                                    & (0.0109)       & (0.0108)                    & (0.0113)                & (1.227)        & (1.243)        & (1.135)\\   
Rent-to-income Ratio  &                & -0.0929$^{**}$              &   &                & -100.5$^{***}$ & \\   
                                    &                & (0.0423)                    &   &                & (17.42)        & \\ 
Tenancy Agreements Pre  &                & -0.0002$^{**}$              &                         &                &                &   \\   
                                    &                & (0.0001)                    &                         &                &                &   \\   
Rent Growth Pre  &                &                             &                         &                & -10.37         &   \\   
                                    &                &                             &                         &                & (6.849)        &   \\
\midrule
Observations & 2,125 & 2,108 & 2,108 & 2,370 & 2,108 & 2,108\\  
Municipalities & 253 & 244 & 244 & 283 & 244 & 244 \\
Time FE &  &  & X &  &  & X \\
Mun FE &  &  & X &  &  & X \\
\bottomrule
\end{tabular}
\begin{tablenotes}
\footnotesize
\item \textit{Notes:} Significance is indicated by \textsuperscript{*} $p < 0.1$, \textsuperscript{**} $p < 0.05$, \textsuperscript{***} $p < 0.01$. Standard errors, in parentheses, are clustered at the municipality level.
\end{tablenotes}
\end{threeparttable}
\end{table}

\subsection{Propensity Score Matching DiD}

To improve the comparability between treated and control municipalities and address concerns regarding the lack of parallel pre-treatment trends in the baseline sample, I re-estimate the DiD regressions using a matched sample obtained through Propensity Score Matching (PSM). I estimate propensity scores using a logistic regression that includes two covariates: the pre-treatment rent growth rate and the rent-to-income ratio. These variables are selected because they align with the criteria used by the regional government to determine whether a municipality qualifies as a stressed residential market area subject to rent control. The treated and control units are matched using a full matching algorithm — a flexible approach that allows for many-to-many matches and retains all observations in the analysis by assigning weights based on the quality of each match.

\begin{table}[htbp]
\centering
\caption{Covariate Balance Before and After Matching}
\label{tab:balance_psm}
\begin{tabular}{lccc}
\toprule
Variable & Diff. Unmatched & Diff. Matched \\
\midrule
Propensity Score & 1.2208 & 0.0019 \\
Rent Price/Net Income & 1.0293 & 0.0134 \\
Rent Growth Pre & 0.4006 & -0.0413 \\
Population & 0.2867 & 0.2876 \\
Gross Income & 0.4238 & 0.2543 \\
Tenancy Agreements Pre & 0.6632 & 1.1813 \\
\bottomrule
\end{tabular}
\end{table}

Table~\ref{tab:balance_psm} presents covariate balance before and after matching. The results indicate that standardized mean differences for the two matching covariates, as well as for the propensity score itself, fall below the conventional threshold of 0.1 after matching — confirming strong balance on these dimensions. The table also reports balance for three additional covariates not used in the matching process. It is encouraging to observe that the differences in other characteristics, such as household income, are also reduced in the matched sample. However, it is worth noting that balance in the number of tenancy agreements per 10,000 inhabitants before the rent control worsens after matching: the standardized difference increases, suggesting that treated and control municipalities are less comparable in this dimension. This limitation, combined with the relatively low effective sample size of the matched control group — only 24 municipalities — highlights the limited overlap in covariate space between treated and control units. Hence, the results of the posterior DiD analysis should be interpreted with some caution.

Table~\ref{tab:psm_did_table_cluster} reports the results of the DiD regressions using the matched sample. Compared to the baseline DiD results, the estimated impact of rent control on rent price growth is considerably smaller and loses statistical significance across all specifications. In the preferred specification with full fixed effects (column 3), the estimated effect is essentially zero (0.0074) and non-significant, suggesting that once better balance is achieved on key covariates, the previously observed reduction in rent growth may not hold. By contrast, the estimated effect on the number of tenancy agreements remains statistically significant and even larger in magnitude than in the unmatched sample. However, this result should be interpreted with caution: as discussed in the previous paragraph, the balance in tenancy agreements was actually worse after matching, indicating that treated and control municipalities became less comparable along this key outcome.

\begin{table}[htbp]
\centering
\caption{Propensity Score Matching DiD Regressions}
\label{tab:psm_did_table_cluster}
\begin{threeparttable}
\begin{tabular}{lcccccc}
\toprule
 & \multicolumn{3}{c}{Rent Growth} & \multicolumn{3}{c}{Tenancy Agreements} \\
 & (1) & (2) & (3) & (4) & (5) & (6) \\
\midrule
Treatment $\times$ Post    & 0.0215        & 0.0204                  & 0.0074        & -8.593$^{***}$ & -4.449$^{*}$   & -6.278$^{***}$\\   
                                    & (0.0430)      & (0.0413)                & (0.0220)      & (3.128)        & (2.559)        & (1.139)\\ 
Rent-to-income Ratio    &               & -0.0984$^{**}$          & 1.086$^{***}$ &                & -107.2$^{***}$ & \\   
                                    &               & (0.0492)                & (0.2645)      &                & (15.76)        & \\
Tenancy Agreements  &               & -0.0007$^{***}$         &               &                &                &   \\   
                                    &               & (0.0003)                &               &                &                &   \\   
Rent Growth  &               &                         &               &                & -8.567         &   \\   
                                    &               &                         &               &                & (6.327)        &   \\  
\midrule
Observations & 2,125 & 2,108 & 2,108 & 2,370 & 2,108 & 2,108\\  
Municipalities & 253 & 244 & 244 & 283 & 244 & 244 \\
Time FE &  &  & X &  &  & X \\
Mun FE &  &  & X &  &  & X \\
\bottomrule
\end{tabular}
\begin{tablenotes}
\footnotesize
\item \textit{Notes:} Significance is indicated by \textsuperscript{*} $p < 0.1$, \textsuperscript{**} $p < 0.05$, \textsuperscript{***} $p < 0.01$. Standard errors, in parentheses, are clustered at the municipality level.
\end{tablenotes}
\end{threeparttable}
\end{table}

\subsection{Synthetic DiD}

Implementing the Synthetic Difference-in-Differences (Synthetic DiD) approach required addressing a key data limitation. This method relies on a complete time series of the outcome variable in the pre-treatment period, meaning that rent growth must be observed for every quarter prior to treatment. As discussed earlier, many control municipalities have missing data in some quarters. In fact, only 27 control municipalities have a fully balanced time series for rent growth—an extremely limited donor pool that risks weakening the validity of the synthetic control.

To address this issue, I employed imputation techniques to fill in missing values of the rent price variable prior to computing rent growth. Specifically, I applied linear interpolation to impute missing values where observations existed both before and after the gap. Once rental prices were fully imputed, rent growth rates were calculated. For the small number of cases with missing data at the start or end of the time series, I used Last Observation Carried Forward (LOCF) and First Observation Carried Backward (FOCB) methods to impute the missing endpoints. While this introduces some assumptions, it greatly expands the number of usable control municipalities and allows us to use the Synthetic DiD methodology as another robustness check.

The results from the Synthetic DiD analysis are broadly consistent with those obtained using Propensity Score Matching. In particular, the estimated effect of rent control on rent growth is again statistically insignificant, with a coefficient of -0.0017. This further reinforces the notion that the decline in rent growth observed in the baseline DiD may be sensitive to model specification and sample composition. However, the analysis continues to show a robust and statistically significant decline in the number of tenancy agreements. Specifically, the results indicate an average reduction of approximately 5 contracts per 10,000 inhabitants in treated municipalities relative to their synthetic counterparts. This provides additional evidence that the rent control policy may have had a negative impact on rental housing supply. However, these results should be interpreted with caution given the limitations of the data and the imputation procedures required for this estimation.

\section{Conclusion}

This paper provides early evidence on the effects of the 2024 rent control policy implemented in Catalonia under the framework of Spain’s national housing law, the \textit{Ley por el Derecho a la Vivienda}. Using newly released administrative data and a difference-in-differences strategy, complemented by event study designs, propensity score matching and synthetic diff-in-diff, I assess the policy’s impact on two key outcomes: rental price growth and the supply of rental housing.

The results suggest that, while the baseline difference-in-differences estimates point to a decline in both rental price growth and the number of tenancy agreements in treated municipalities, the validity of these findings is called into question by the evidence against the parallel trends assumption. When using alternative identification strategies that improve the comparability between treated and control municipalities, the estimated effect on rent growth becomes statistically insignificant. However, the decline in tenancy agreements remains robust across specifications, suggesting a reduction of the number of tenancy agreements by around 5 - 7 per 10,000 inhabitants.

These conclusions should be interpreted with caution due to several data limitations. First, the rental price variable is only available as an unadjusted municipal average, which may introduce composition effects related to housing size or quality. Second, a significant share of control municipalities present many missing values, leading to a smaller and potentially non-representative control pool. Lastly, the most recent data only extends through 2024, limiting post-treatment observations, particularly for municipalities treated later in the year.

Despite these limitations, this study offers a first empirical look at a newly implemented policy and raises important questions about the effectiveness and unintended consequences of rent control. Future research should revisit this analysis as more data becomes available, particularly for the second wave of treated municipalities, and explore additional outcomes such as shifts to tourist rentals, housing ownership or changes in housing quality. Such work is essential to guide evidence-based policymaking in the area of housing affordability.

\nocite{*}

\clearpage

\bibliography{citations}

@article{sims2007,
  title={Out of control: What can we learn from the end of Massachusetts rent control?},
  author={Sims, David P},
  journal={Journal of Urban Economics},
  volume={61},
  number={1},
  pages={129--151},
  year={2007},
  publisher={Elsevier}
}

@article{autor2014,
  title={Housing market spillovers: Evidence from the end of rent control in Cambridge, Massachusetts},
  author={Autor, David H and Palmer, Christopher J and Pathak, Parag A},
  journal={Journal of Political Economy},
  volume={122},
  number={3},
  pages={661--717},
  year={2014},
  publisher={University of Chicago Press Chicago, IL}
}

@article{diamond2019,
Author = {Diamond, Rebecca and McQuade, Tim and Qian, Franklin},
Title = {The Effects of Rent Control Expansion on Tenants, Landlords, and Inequality: Evidence from San Francisco},
Journal = {American Economic Review},
Volume = {109},
Number = {9},
Year = {2019},
Month = {September},
Pages = {3365–94}}

@article{mense2019,
Author = {Mense, Andreas and Michelsen, Claus and Kholodilin, Konstantin A.},
Title = {The Effects of Second-Generation Rent Control on Land Values},
Journal = {AEA Papers and Proceedings},
Volume = {109},
Year = {2019},
Month = {May},
Pages = {385–88}}

@article{mense2023,
  title={Rent control, market segmentation, and misallocation: Causal evidence from a large-scale policy intervention},
  author={Mense, Andreas and Michelsen, Claus and Kholodilin, Konstantin A},
  journal={Journal of Urban Economics},
  volume={134},
  pages={103513},
  year={2023},
  publisher={Elsevier}
}

@article{breidenbach2022,
  title={Temporal dynamics of rent regulations--The case of the German rent control},
  author={Breidenbach, Philipp and Eilers, Lea and Fries, Jan},
  journal={Regional Science and Urban Economics},
  volume={92},
  pages={103737},
  year={2022},
  publisher={Elsevier}
}

@book{monras2022,
  title={The effect of second generation rent controls: New evidence from Catalonia},
  author={Monras, Joan and Montalvo, Jos{\'e} Garcia and others},
  year={2022},
  publisher={Universitat Pompeu Fabra, Department of Economics and Business}
}

@article{jofre2023,
  title={Effectiveness and supply effects of high-coverage rent control policies},
  author={Jofre-Monseny, Jordi and Mart{\'\i}nez-Mazza, Rodrigo and Seg{\'u}, Mariona},
  journal={Regional Science and Urban Economics},
  volume={101},
  pages={103916},
  year={2023},
  publisher={Elsevier}
}

@article{arkhangelsky2021,
  title={Synthetic difference-in-differences},
  author={Arkhangelsky, Dmitry and Athey, Susan and Hirshberg, David A and Imbens, Guido W and Wager, Stefan},
  journal={American Economic Review},
  volume={111},
  number={12},
  pages={4088--4118},
  year={2021},
  publisher={American Economic Association 2014 Broadway, Suite 305, Nashville, TN 37203}
}

@article{glaeser2003,
  title={The misallocation of housing under rent control},
  author={Glaeser, Edward L and Luttmer, Erzo F P},
  journal={American economic review},
  volume={93},
  number={4},
  pages={1027--1046},
  year={2003},
  publisher={American Economic Association}
}

@article{favilukis2023,
  title={Affordable housing and city welfare},
  author={Favilukis, Jack and Mabille, Pierre and Van Nieuwerburgh, Stijn},
  journal={The Review of Economic Studies},
  volume={90},
  number={1},
  pages={293--330},
  year={2023},
  publisher={Oxford University Press}
}
\bibliographystyle{apacite}

\end{document}